\documentclass[twocolumn,aps,superscriptaddress,prb,longbibliography]{revtex4-1}
\setcitestyle{numbers,square}
\usepackage{amsmath,amssymb}
\usepackage{tabularx}
\usepackage{graphicx}
\usepackage{bmpsize}
\usepackage{bm}
\usepackage{color}
\usepackage{amssymb}
\usepackage[version=3]{mhchem}
\usepackage{newtxmath}
\usepackage{physics}
\usepackage{hyperref}
\DeclareMathAlphabet{\mathpzc}{OT1}{pzc}{m}{it}

\begin{document}
\title{
Topological gap labeling with the third Chern numbers in three-dimensional quasicrystals
}
\author{Kazuki Yamamoto}
\affiliation{Department of Physics, Osaka University, Toyonaka, Osaka 560-0043, Japan}
\author{Mikito Koshino}
\affiliation{Department of Physics, Osaka University, Toyonaka, Osaka 560-0043, Japan}
\date{\today}

\begin{abstract}
We study the topological gap labeling of general 3D quasicrystals
and we find that every gap in the spectrum is characterized by a set of the third Chern numbers.
We show that a quasi-periodic structure has multiple Brillouin zones defined by redundant wavevectors, 
and the number of states below a gap is quantized as an integer linear combination of 
volumes of these Brillouin zones.
The associated quantum numbers to characterize energy gaps can be expressed as third Chern numbers
by considering a formal relationship between an adiabatic charge pumping under cyclic deformation 
of the quasi-periodic potential and a topological nonlinear electromagnetic response in 6D insulators.
\end{abstract}
\maketitle
\section{introduction}
\label{sec:intro}

Quasicrystals are non-periodic but long-range ordered systems found in a wide variety of physical systems including metallic alloys\cite{Shechtman1984,Elser1985,Levine1984,Kamiya2018}, photonic quasicrystals\cite{Vardeny2013,Sanchez-Palencia2005,Kraus2012,Bandres2016}, ultra cold-atom systems  \cite{Nakajima2021,Price2015,Price2016}  and twisted two-dimensional (2D) materials. \cite{Fujimoto2021,Oka2021,Su2020,Yinhan2020,Fujimoto2020}
Despite the increasing importance of quasicrystalline systems, the theoretical description of their physical properties is limited by the lack of the Bloch theorem. In periodic crystals, the energy spectrum is quantized into the Bloch bands with equal numbers of states,
which corresponds to the area of the Brillouin zone (BZ).
Therefore each energy gap is characterized by an integer, which is the number of the bands below the gap.
In contrast, it is supposed that quasicrystals do not have such a quantum unit to count the number of states, but rather the spectrum splits to a set of infinitely many bands (the Cantor set) as 
the infinite-period limit of a periodic system.

In our previous works \cite{koshino2021,Oka2021}, we studied spectral quantization of general 2D quasi-periodic systems and showed that the gap labeling is actually possible in the following sense.
Specifically, the energy spectrum of a quasicrystal is characterized by multiple 
BZs defined with redundant wavevectors, and the number of states below the gap is always quantized as an integer linear combination of the areas of these BZs. 
The quantum numbers to characterize energy gaps were shown to be topological invariants expressed as the second Chern numbers, by considering a mapping between 2D quasicrystals and four-dimensional quantum Hall insulators.
Topological characterization of energy gaps in quasicrystals was also studied in different contexts
for in one-dimensional (1D)\cite{lang2012edge,mei2012simulating,kraus2012topological,kraus2012topologicalequivalence,satija2013chern,ganeshan2013topological,verbin2013observation,verbin2015topological,lohse2016thouless,marra2020topologically,zilberberg2021topology,yoshii2021charge} and two-dimensional (2D) quasiperiodic systems
 \cite{kraus2013four,tran2015topological,bandres2016topological,cain2020layer,rosa2021topological,fujimoto2020topological,zhang2020topological,su2020topological},
while the gap labeling of three dimensional (3D) quasicrystals is yet to be explored.

\begin{figure}
\centering
\includegraphics[width=0.9\linewidth]{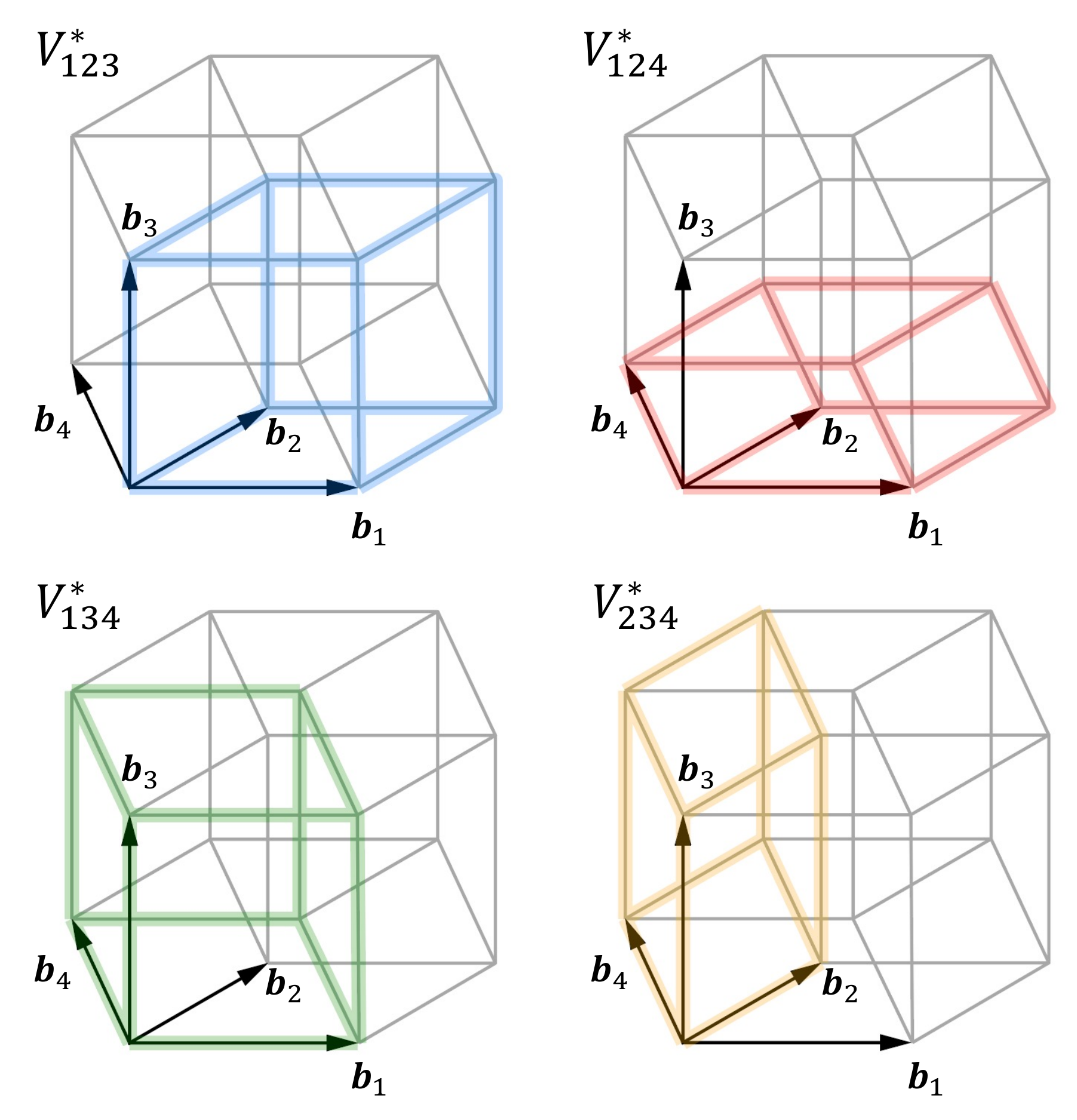}
\caption{Fundamental Brillouin zones in a 3D quasiperiodic system with four reciprocal lattice vectors,
$\bm{b}_1,\bm{b}_2,\bm{b}_3,\bm{b}_4$}
\label{fig_multiple_BZs}
\end{figure}

In this paper, we extend the argument for 2D  \cite{koshino2021,Oka2021} to 3D,
and show that the spectrum of a 3D quasicrystal is quantized by the third Chern numbers, which correspond to 
electromagnetic response in six-dimensional (6D) insulator.
We consider a general 3D quasicrystalline system 
with the number of reciprocal lattice vectors greater than the number of the spatial dimensions.
Specifically, it is described by the Hamiltonian in a 3D space, 
\begin{align}
&H =\frac{\bm{p}^2}{2m} + U(\bm{r}),
\notag\\
&U(\bm{r})=\sum_{m_1,...,m_N}U_{m_1,...,m_N}e^{i\sum_{i=1}^{N}m_i\bm{b}_i\cdot\bm{r}},
    \label{eq_V}
\end{align}
where $m_1,...,m_N$ are integers and $\bm{b}_i\,(i=1,2,...,N)$ are a set of redundant reciprocal lattice vectors ($N>3$). 
By taking three distinct vectors from the set, 
we can define fundamental BZs with volume of $V_{ijk}^{*}=\bm{b}_i\cdot(\bm{b}_j \times \bm{b}_k)$,
as illustrated in Fig.~\ref{fig_multiple_BZs} for the case of $N=4$.
We claim that, when the energy spectrum has a gap, 
the electron density below the gap is quantized as
\begin{equation}
    n_e=\frac{1}{(2\pi)^3}\sum_{ijk}C_{ijk}^{(3)}V_{ijk}^{*},                                               \label{eq1}
\end{equation}
where $C_{ijk}^{(3)}$ is the third Chern number 
calculated from the occupied states.
As we have $N!/[(N-3)!3!]$ choices of $(i,j,k)$,
every single gap is characterized by a set of $N!/[(N-3)!3!]$ third Chern numbers.
The statement can be proved by considering a formal relationship between
an adiabatic charge pumping under cyclic deformation of the potential
and a topological electromagnetic response in a fictitious 6D insulator \cite{petrides2018six}.


This paper is organized as follows. In Sec.~\ref{sec_b2}, we present a general description of the electromagnetic response of the (6+1)D system using an effective action formalism.
In Sec.~\ref{sec_b3}, we consider an adiabatic pumping 
in the 3D quasicrystal and a mapping to the 6D system. With the aid of the formula obtained in Sec.~\ref{sec_b2} and the dimensional reduction technique, we will finally obtain the result Eq.~\eqref{eq1}.
A brief conclusion is given in Sec.~\ref{sec_b4}.
Throughout the paper, we use the natural unit $\hbar=c=e=1$ and the Minkowski metric $\eta_{\mu\nu}=\rm{diag}(-1,+1,...,+1)$.
\section{Electromagnetic response of (6+1)D systems}
\label{sec_b2}
In this section, we describe a topological nonlinear response of a generic 6D band insulator in an electromagnetic field, and express the response coefficient with a third Chern number.
The problem was also studied by the semiclassical approach. \cite{petrides2018six,lee2018electromagnetic}
Here we use the Euclidean path integral formalism, by extending the argument for (4+1)D systems \cite{Qi2008}
to (6+1)D. 
The effective action $S_{\rm{eff}}$ in (6+1)D is defined as 
\begin{equation}
    e^{iS_{\rm{eff}}}=\int Dc^{\dag}Dc \, e^{-S-S_{\rm{int}}}
    \label{eq_e_S_eff}
\end{equation}
where
\begin{align}
    S&=\int d\tau\sum_{\bm{k}}c^{\dag}_{\bm{k}}(\tau)\Bigl(\pdv{\tau}-h(\bm{k})\Bigr) c_{\bm{k}}(\tau) \\
    &=\sum_{\bm{k}n}c^{\dag}_{\bm{k}n}\Bigl(i\omega_n-h(\bm{k})\Bigr) c_{\bm{k}n},\\
    S_{\rm{int}}&=\sum_{\bm{q}\omega} A^\mu(\bm{q},\omega) J_\mu(-\bm{q},-\omega).
\end{align}
Here, $\tau=-it$ is imaginary time and
$A^\mu(\bm{q},\omega)=(A^0,\bm{A})$ is an external electromagnetic four-potential with wavenumber $\bm{q}$ and frequency $\omega$. The one-particle Hamiltonian is represented by
$h(\bm{k})$, and $c^{\dag}_{\bm{k}n}$ and $c_{\bm{k}n}$ are Grassmann numbers of an electron with Bloch wavenumber $\bm{k}$ and Matsubara frequency $\omega_n$. The current $J^\mu=(J^0,\bm{J})$ is expressed as
\begin{align}
    J^0(\bm{q},\omega)&=-\sum_{\bm{k}n} c^{\dag}_{\bm{k}+\bm{q},\omega_n+\omega}c_{\bm{k},\omega_n}\\
    \bm{J}(\bm{q},\omega)&=-\sum_{\bm{k}n} \nabla_{\bm{k}} h(\bm{k}) c^{\dag}_{\bm{k}+\bm{q},\omega_n+\omega}c_{\bm{k},\omega_n}.
\end{align}

For the (6+1)-dimensional insulator, the effective action contains a topological term called the third Chern-Simons term,
\begin{equation}
    S_{\rm{eff}}=\frac{C^{(3)}}{192\pi^3}\int A\wedge dA\wedge dA\wedge dA,                                                      \label{eq7}
\end{equation}
with
\begin{align}
    C^{(3)}=\frac{\pi^3}{105}\int\frac{d^7l}{(2\pi)^7}\epsilon^{\mu\nu\rho\sigma\tau\lambda\delta}\tr G(\partial_{\mu}G^{-1})G(\partial_{\nu}G^{-1})\notag\\ 
    \times G(\partial_{\rho}G^{-1})G(\partial_{\sigma}G^{-1})G(\partial_{\tau}G^{-1})G(\partial_{\lambda}G^{-1})G(\partial_{\delta}G^{-1}),                 \label{eq2}
\end{align}
where $l^{\mu}=(i\omega,\bm{l})$ is the frequency-momentum vector and $G(l)=(i\omega-h(\bm{l}))^{-1}$ is the one-particle Green's function. 
The detailed derivation of Eqs.\~(\ref{eq7}) and (\ref{eq2}) is presented in Appendix \ref{sec_app1}.
We obtain the topological nonlinear response to an external electromagnetic field $A^\mu$ as
\begin{align}
    j^\mu&=\fdv{S_{\rm{eff}}}{A_\mu}\notag \\
    &=\frac{C^{(3)}}{48\pi^3}\epsilon^{\mu\nu\rho\sigma\tau\lambda\delta}\partial_{\nu}A_{\rho}\partial_{\sigma}A_{\tau}\partial_{\lambda}A_{\delta}.\label{eq4}
\end{align}

The coefficient $C^{(3)}$ in Eq.~(\ref{eq2}) is expressed as the third Chern number 
of the non-abelian Berry connection in the 6D Brillouin zone (BZ). Specifically, it is written as
\begin{equation}
    C^{(3)}=\frac{1}{48\pi^3}\int_{BZ} \tr f\wedge f\wedge f,                                                                      \label{eq3}
\end{equation}
where $f$ is the Berry curvature defined by
\begin{align}
f &=\frac{1}{2}f_{ij}\, dk^i\wedge dk^j
\notag\\
f_{ij}^{\alpha\beta}  &=\partial_ia_j^{\alpha\beta}-\partial_ja_i^{\alpha\beta}-i[a_i,a_j]^{\alpha\beta}, \notag\\
a_i^{\alpha\beta} 
&=i\bra{\alpha\bm{k}}\pdv{k^i}\ket{\beta\bm{k}},
\end{align}
and the indices $\alpha,\beta$ represent the occupied bands. The derivation of Eq.~(\ref{eq3}) is described in Appendix \ref{sec_app2}.
The $C^{(3)}$ is a topological number which is invariant under continuous deformations without closing an energy gap. 

\section{Topological numbers in 3D quasi-periodic systems}
\label{sec_b3}
\subsection{Adiabatic quantum pumping}

Let us consider a 3D quasicrystalline system expressed by Eq.~(\ref{eq_V}),
and calculate the adiabatic charge pumping under a cyclic change of the potential $U(\bm{r})$.
We introduce phase parameters $\phi_1,...,\phi_N$ to the potential as
\begin{align}
    U(\bm{r};\,\phi_1,...,\phi_N) =&\sum_{m_1,...m_N}U_{m_1,...,m_N}e^{i\sum_{i=1}^{N}m_i(\bm{b}_i\cdot\bm{r}-\phi_i)},
    \label{eq_V_phi}
\end{align}
and consider a cyclic process where $\phi_i$, with a certain $i$, is adiabatically increased from 0 to $2\pi$. 
In a periodic case with $N=3$, the process corresponds to just a parallel translation of the potential $U(\bm{r})$ by a real-space lattice period $\bm{a}_i$ where $\bm{a}_i\cdot\bm{b}_j=2\pi\delta_{ij}$.
Eq.~\eqref{eq_V_phi} is a generalization to quasiperiodic systems,
while it is not generally expressed as a simple translation.
If the potential $U(\bm{r})$ is a summation of
independent periodic potentials
$U_1(\bm{r}), U_2(\bm{r}), \cdots$ not sharing the same $\bm{b}_i$, 
in particular,
a change of $\phi_i$ is equivalent to a relative sliding of a $U_n$
with respect to the rest $U_m$'s.
In 2D, this corresponds to interlayer sliding in moir\'{e} multilayer systems.
\cite{Fujimoto2020,zhang2020topological,su2020topological,Oka2021}

We define $\Delta \bm{P_i}$ as the change of the electric polarization during a
single cycle from $\phi_i=0$ to $2\pi$.
When the spectrum has an energy gap, $\Delta \bm{P_i}$ is given by
\begin{equation}
    \Delta \bm{P_i}=2\pi\pdv{n_e}{\bm{b}_i},                                                               \label{eq5}
\end{equation}
where $n_e$ is the electron density below the energy gap.\cite{koshino2021} Eq.~\eqref{eq5} can be proved by the following consideration.
When a specific reciprocal lattice vector $\bm{b}_i$ is infinitesimally changed to $\bm{b}_i+\delta\bm{b}_i$, this leads to a change to the potential $V$ at a point $\bm{r}$ which is equivalent to a phase change by $\delta \phi_i=-\delta\bm{b}_i\cdot\bm{r}$. This causes a polarization change by
\begin{equation}
    \Delta \bm{P}_i\frac{\delta\phi_i}{2\pi}=\Delta \bm{P_i}\frac{(-\delta \bm{b}_i\cdot\bm{r})}{2\pi}.
\end{equation}
Now, we consider a closed curved surface $S$, and let $N_e$ be the number of electrons inside $S$. When $\bm{b}_i$ is changed to $\bm{b}_i+\delta\bm{b}_i$, the number of electrons passing through $S$ is
\begin{align}
    \delta N_e &=\int_S  \frac{\delta \bm{b}_i\cdot\bm{r}}{2\pi}\Delta \bm{P_i}\cdot  d\bm{S}
    \notag \\
    &=\int_\Omega  {\rm div} \left(\frac{\delta \bm{b}_i\cdot\bm{r}}{2\pi}\Delta \bm{P_i}\right)
    dV
    =\frac{\Omega}{2\pi}\Delta \bm{P_i}\cdot\delta \bm{b}_i,
\end{align}
where 
$\Omega$ is the volume enclosed inside of $S$. Noting that the electron density is defined as $n_e=N_e/\Omega$, we obtain Eq.~\eqref{eq5}.

\subsection{Mapping to a $(3+N)$-dimensional system}

The adiabatic charge pumping in 3D quasicrystal discussed above can be described 
in an alternative approach considering an electromagnetic magnetic response in a ($3+N$)-dimensional system.
By using the mapping, we will show that the transferred charge 
in the pumping is interpreted as integer-quantized response current in 6D\cite{petrides2018six},
and it finally leads to the zone quantization rule, Eq.~\eqref{eq1}.
The formulation is basically an extension of the argument for 2D quasicrystal \cite{koshino2021} to 3D.

We consider a ($3+N$)D system in $(x,y,z,w_1,w_2,...,w_N)$ space, which is continuous in $x,y,$ and $z$ directions and discrete in $w_i(i=1,2,...,N)$ directions with lattice spacing $a_i$. For the $w_i$-direction, we assume nearest-neighbor tight-binding coupling $t_i$ between adjacent layers. We apply a uniform magnetic field $B_{xi}$, $B_{yi}$ and $B_{zi}$ perpendicular to $xw_i$-plane, $yw_i$-plane and $zw_i$-plane, respectively. We take the vector potential as $\bm{A}=\sum_{i=1}^{N} (B_{xi}x+B_{yi}y+B_{zi}z)\bm{e}_i$, where $\bm{e}_i$ is the unit vector in the $w_i$-direction. 

Since the Hamiltonian is periodic in any of the $w_i$-directions, the wavefunction can be written as $\Psi(x,y,z,w_1,w_2,...,w_N)=\psi(x,y,z)e^{i\sum_{i}k_iw_i}$, where $k_i$ is the Bloch wave number defined in $-\pi/a_i\leq k_i \leq \pi/a_i$. The ($3+N$)D Schr{\"o}dinger equation is reduced to the 3D equation as
\begin{equation}
    \frac{\bm{p}^2}{2m}\psi(x,y,z)-\sum_{i=1}^{N}2t_i \cos{(\bm{b}_i\cdot\bm{r}+\phi_i)}=E\psi(x,y,z),
    \label{eq_3D}
\end{equation}
where 
\begin{align}
&    \bm{b}_i=a_i\bm{B}_i=a_i(B_{xi},B_{yi},B_{zi}), \\
&    \phi_i=k_ia_i. 
\end{align}
This is nothing but a 3D quasi-periodic system considered in the previous seciton. 
Higher harmonic terms in $\bm{b}_i$ can be incorporated by including further-range hoppings in $w_i$ direction in the original $(3+N)$D  model.

Now we consider an electronic response of the $(3+N)$D system to a weak external electric field $E_i$ applied in the $w_i$ direction.
The $E_i$ adiabatically changes the wavenumber $k_i$ as $dk_i/dt = -E_i$,
where the factor $-1$ is the charge of an electron in natural unit.
In the corresponding 3D equation, Eq.~(\ref{eq_3D}),
it is equivalent to an adiabatic potential change by shifting $\phi_i$,
which was considered in the previous section.
A cyclic change from $\phi_i=0$ to $2\pi$ corresponds to
a translation of $k_i$ by the Brillouin zone width, ${2\pi}/{a_i}$,
which takes a time $T = (2\pi /a_i)/E_i$. 

We assume that the Fermi energy is in an energy gap in the $(3+N)$D  system.
The response electric current induced by $E_i$
is obtained by calculating those for 6D subspaces $(x,y,z,w_i,w_j,w_k)$,
and taking a sum over indeces $j,k(\neq i)$.
According to Eq.~\eqref{eq4},
the response current in the 6D subspace is given by 
\begin{equation}
    \bm{j}^{(6D)}=\frac{C_{ijk}^{(3)}}{8\pi^3}E_i(\bm{B}_j \times \bm{B}_k).
\end{equation}
The corresponding 3D current density per layer is given by $\bm{j}^{(3D)}=\bm{j}^{(6D)}a_i a_j a_k$, leading to
\begin{equation}
    \bm{j}^{(3D)}=\frac{C_{ijk}^{(3)}}{(2\pi)^2T}\bm{b}_j \times \bm{b}_k.
\end{equation}
The total polarization change in a cyclic process is $\Delta \bm{P}_i=\bm{j}^{(3D)}T$. Taking summations over $j$ and $k$, we obtain
\begin{equation}
    \Delta \bm{P}_i=\frac{1}{(2\pi)^2}\sum_{jk}C_{ijk}^{(3)}\bm{b}_j \times \bm{b}_k.                       \label{eq6}
\end{equation}
By applying Eq.~\eqref{eq6} to Eq.~\eqref{eq5}, we finially obtain the result 
\begin{equation}
n_e=\frac{1}{(2\pi)^3}\sum_{ijk}C_{ijk}^{(3)}
\bm{b}_i\cdot(\bm{b}_j \times \bm{b}_k),
\end{equation}
which is Eq.~\eqref{eq1}. 

The result is analogous to 2D quasicrystal where $n_e$
is quantized by the second Chern number \cite{koshino2021},
and also to 1D quasicrystal quantized by the first Chern number \cite{Kraus2012}.
The calculation for the Chern numbers requires the Brillouin zone,
and practically it can be achieved by considering 
a commensurate approximant,\cite{koshino2021,Oka2021} where the periodicities of $\bm{b}_i(i=1,2,...,N)$ have a common super unit cell.


\section{conclusion}
\label{sec_b4}
We have provided a topological concept to characterize energy gaps in 3D quasicrystals. We found that the electron density below the gap is quantized as an integer linear combination of volumes of multiple Brillouin zones, which are defined by redundant reciprocal lattice vectors in the quasi-periodic system. Then we showed that these integers can be expressed as
the third Chern numbers by considering a mapping between
the 3D quasicrystal and a $(3+N)$D band insulator.
Specifically, an adiabatic charge pumping in a potential phase change in the 3D system (a generalization of a relative slide of multiple periodic potentials)
can be viewed as a projection of the nonlinear electromagnetic response in 6D subspaces in $(3+N)$D system,
and the latter is shown to be described by the third Chern numbers.
The gap characterization scheme presented in this work is applicable to general 3D quasicrystalline systems having redundant periodicities more than the number of the spatial dimensions.


\section*{Acknowledgments}

This work was supported in part by 
JSPS KAKENHI Grant Number JP20H01840, JP20H00127, JP21H05236, JP21H05232 and by JST CREST Grant Number JPMJCR20T3, Japan.

\appendix
\begin{widetext}
\section{Derivation of Eq.~\eqref{eq7} and (\ref{eq2}) }
\label{sec_app1}

Here we show that the effective action $S_{\rm eff}$
of a 6D band insulator under an eletromagnetic field  [Eq.~\eqref{eq_e_S_eff}]
includes the term of Eq.~\eqref{eq7} with Eq.~(\ref{eq2}).
We concentrate on the term proportional to $A^4$ in $S_{\rm eff}$, and define the four-point function $\Pi^{\mu\rho\tau\delta}(x,y,z,w)$ as
\begin{equation}
    S_{\rm{eff}}=\frac{1}{4!}\int d^7x \int d^7y \int d^7z \int d^7w\Pi^{\mu\rho\tau\delta}(x,y,z,w)A_\mu(x)A_\rho(y)A_\tau(z)A_\delta(w)\notag.
\end{equation}
Then $\Pi^{\mu\rho\tau\delta}(x,y,z,w)$ can be represented by
\begin{equation}
     i\Pi_{\mu\rho\tau\delta}(x,y,z,w)=\frac{\int Dc^{\dag}Dc \, J_{\mu}(x)J_{\rho}(y)J_{\tau}(z)J_{\delta}(w) \, e^{-S} }{\int Dc^{\dag}Dc \, e^{-S}}\notag.
\end{equation}
Since the current $J_{\mu}(x)$ satisfies the continuity equation $\partial_\mu J^{\mu}=0$, $\Pi^{\mu\rho\tau\delta}(x,y,z,w)$ must satisfy
\begin{equation}
    \pdv{x^\mu}\Pi^{\mu\rho\tau\delta}(x,y,z,w)
    =\pdv{y^\rho}\Pi^{\mu\rho\tau\delta}(x,y,z,w)
    =\pdv{z^\tau}\Pi^{\mu\rho\tau\delta}(x,y,z,w)
    =\pdv{w^\delta}\Pi^{\mu\rho\tau\delta}(x,y,z,w)
    =0.\notag
\end{equation}
This requirement suggests that the term,
\begin{equation}
    \Pi^{\mu\rho\tau\delta}(x,y,z,w)=s\epsilon^{\mu\nu\rho\sigma\tau\lambda\delta}\pdv{y^\nu}\delta(y-x)\pdv{z^\sigma}\delta(z-x)\pdv{w^\lambda}\delta(w-x)+\cdots \notag
\end{equation}
should be included in $\Pi^{\mu\rho\tau\delta}(x,y,z,w)$,
where $s$ is a certain constant.
This term is specific to the $(6+1)$D system as it has 7 indices.
Taking the Fourier transform of $\Pi^{\mu\rho\tau\delta}(x,y,z,w)$, we obtain
\begin{align}
    \Pi^{\mu\rho\tau\delta}(r,p,q,k)&=\int d^7x e^{-irx} \int d^7y e^{-ipy} \int d^7z e^{-iqz}\int e^{-ikw} \Pi^{\mu\rho\tau\delta}(x,y,z,w)\notag\\
    &=(2\pi)^7\delta(r+p+q+k)\Tilde{\Pi}^{\mu\rho\tau\delta}(p,q,k),
\end{align}
where
\begin{align}
   \Tilde{\Pi}^{\mu\rho\tau\delta}(p,q,k)
     =-is\epsilon^{\mu\nu\rho\sigma\tau\lambda\delta}p_\nu q_\sigma 
     k_\lambda+\cdots.
\end{align}
The constant $s$ is given by
\begin{equation}
 s=\eval{\frac{1}{7!}\epsilon^{\mu\nu\rho\sigma\tau\lambda\delta}\pdv{p^\nu}\pdv{q^\sigma}\pdv{k^\lambda}i\Tilde{\Pi}_{\mu\rho\tau\delta}(p,q,k)}_{p=q=k=0}.        \label{eq8}
\end{equation}

\begin{figure}[t]
    \centering
    \includegraphics[scale=0.7]{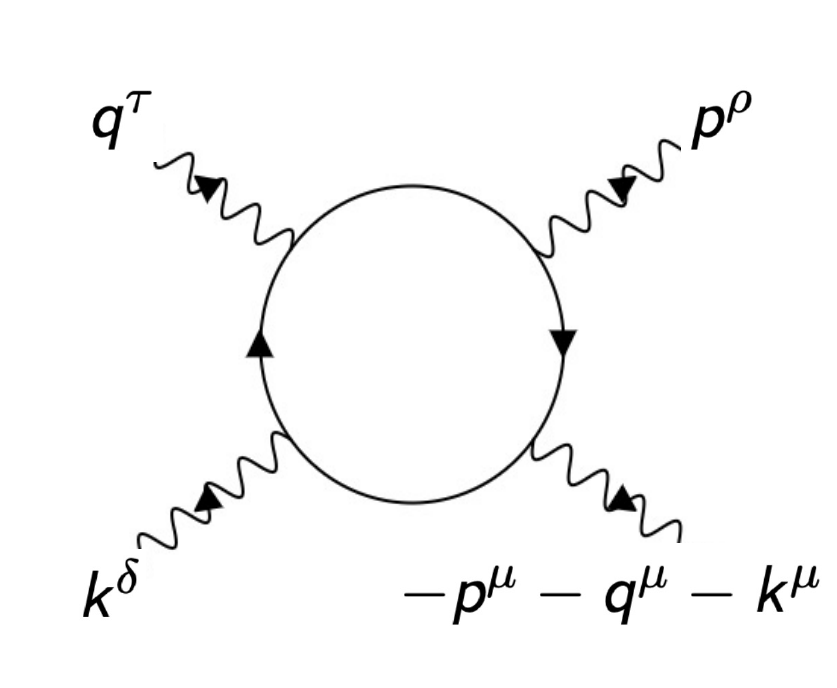}
    \caption{One-loop Feynman diagram that contributes to $F_{\delta\tau\rho\mu}(k,q,p)$. The solid and wavy lines correspond to electron and the external electromagnetic field, respectively.}
    \label{fig:1}
\end{figure}

The four point function $i\Tilde{\Pi}^{\mu\rho\tau\delta}(p,q,k)$ has contributions from $3!$ Feynman diagrams. One of them is illustrated in Fig. \ref{fig:1}, and others are obtained by permutation. We can explicitly perform path integrals, giving
\begin{align}
    &i\Tilde{\Pi}_{\mu\rho\tau\delta}(p,q,k)\notag\\
    =&\frac{\int Dc^{\dag}Dc \, J_{\mu}(-p-q-k)J_{\rho}(p)J_{\tau}(q)J_{\delta}(k) \, e^{-S} }{\int Dc^{\dag}Dc \, e^{-S}}\notag\\
    =&F_{\delta\tau\rho\mu}(k,q,p)+F_{\delta\rho\tau\mu}(k,p,q)+F_{\tau\delta\rho\mu}(q,k,p)+F_{\tau\rho\delta\mu}(q,p,k)+F_{\rho\tau\delta\mu}(p,q,k)+F_{\rho\delta\tau\mu}(p,k,q),\label{eq9}
\end{align}
where we define 
\begin{equation}
    F_{\delta\tau\rho\mu}(k,q,p)=-\int\frac{d^7l}{(2\pi)^7}\tr G(l)\pdv{G^{-1}(l)}{l^\delta} G(l+k)\pdv{G^{-1}(l+k)}{l^\tau}G(l+k+q)\pdv{G^{-1}(l+k+q)}{l^\rho}G(l+k+q+p)\pdv{G^{-1}(l)}{l^\mu},\notag
\end{equation}
and the minus sign originates from the fermion loop.
In the calculation, we used the expression
\begin{equation}
    J^\mu(\bm{q},\omega)=\sum_{k,n}c^{\dag}_{\bm{k}+\bm{q},\omega_n+\omega}\pdv{G^{-1}}{k_\mu}c_{\bm{k},\omega_n}.\notag
\end{equation}
By applying Eq.~\eqref{eq9} to Eq.~\eqref{eq8}, we finally obtain
\begin{equation}
    s=-\frac{3!}{7!}\int\frac{d^7l}{(2\pi)^7}\epsilon^{\mu\nu\rho\sigma\tau\lambda\delta}\tr G(\partial_{\mu}G^{-1})G(\partial_{\nu}G^{-1})G(\partial_{\rho}G^{-1})G(\partial_{\sigma}G^{-1})G(\partial_{\tau}G^{-1})G(\partial_{\lambda}G^{-1})G(\partial_{\delta}G^{-1}) \notag.
\end{equation}

\section{Derivation of Eq.~\eqref{eq3}}
\label{sec_app2}

Let us show that the coefficient $C^{(3)}$ in Eq.~(\ref{eq2}) is expressed as the third Chern number 
as in Eq.~\eqref{eq3}.
The derivation is closely analogous to Ref.~\cite{Qi2008}, which investigated the classification of (4+1)D time reversal invariant topological insulators in terms of the 2nd Chern number and (4+1)D Chern-Simons theory. Here we extend the argument to (6+1)D.

First, we show that any continuous deformation of $h(\bm{k})$ does not change Eq.~\eqref{eq2}. When $h(\bm{k})$ is infinitesimally changed to $h(\bm{k})+\delta h$, the Green's function $G$ is changed to $G+\delta G$. The change in each factor $G(\partial_{\mu}G^{-1})$ in Eq.~\eqref{eq2} makes the same contribution to the change in Eq.~\eqref{eq2}, giving
\begin{equation}
    \delta C^{(3)}=7\times\frac{\pi^3}{105}\int\frac{d^7l}{(2\pi)^7}\epsilon^{\mu\nu\rho\sigma\tau\lambda\delta}\tr \delta\Bigl(G(\partial_{\mu}G^{-1})\Bigr)G(\partial_{\nu}G^{-1})G(\partial_{\rho}G^{-1})G(\partial_{\sigma}G^{-1})G(\partial_{\tau}G^{-1})G(\partial_{\lambda}G^{-1})G(\partial_{\delta}G^{-1}),\notag
\end{equation}
where the change of the factor $G(\partial_{\mu}G^{-1})$ is given by
\begin{equation}
    \delta\Bigl(G(\partial_{\mu}G^{-1})\Bigr)=-G\partial_\mu(G^{-1}\delta G)G^{-1}.\notag
\end{equation}
Integrating by parts, we obtain $\delta C^{(3)}=0$. 

Without loss of generality, the chemical potential can be defined to be zero. Since any gapped Hamiltonian $h(\bm{k})$ can be continuously deformed into the simple Hamiltonian $h_0(\bm{k})$, that is the form
\begin{align}
    h_0(\bm{k})&=\epsilon_G\sum_{1\leq \alpha \leq M} \ket{\alpha\bm{k}}\bra{\alpha \bm{k}}+\epsilon_E \sum_{M+1\leq \alpha'} \ket{\alpha'\bm{k}}\bra{\alpha' \bm{k}}\notag\\
    &=\epsilon_GP_G(\bm{k})+\epsilon_EP_E(\bm{k}),\notag
\end{align}
where $P_G (P_E)$ is the projection operator of ground states (excited states), $\alpha=1,2,...M$ are occupied bands, and $\alpha '=M+1,...$ are unoccupied bands. Here, $\epsilon_G (\epsilon_E)$ is the energy of the ground states (excited states) and satisfies $\epsilon_G<0<\epsilon_E$. Therefore, it is sufficient to prove Eq.~\eqref{eq3} for the simple Hamiltonian $h_0(\bm{k})$. In this case, the one-particle Green's function is written as
\begin{align}
    G(\bm{k},\omega)&=\frac{1}{i\omega-\epsilon_G P_G(\bm{k})-\epsilon_E P_E(\bm{k})}
    =\frac{P_G(\bm{k})}{i\omega-\epsilon_G}+\frac{P_E(\bm{k})}{i\omega-\epsilon_E}.\notag\\
\end{align}
The derivatives of $G^{-1}(\bm{k},\omega)$ are calculated as
\begin{align}
    \pdv{G^{-1}}{k^0}\qty(\bm{k},\omega)&=1,\notag\\
    \pdv{G^{-1}}{k^i}\qty(\bm{k},\omega)&=-\epsilon_G\pdv{P_G}{k^i}\qty(\bm{k})-\epsilon_E\pdv{P_E}{k^i}\qty(\bm{k})
    =(\epsilon_E-\epsilon_G)\pdv{P_G}{k^i}\qty(\bm{k}), \quad(i=1,2,3,4,5,6).\notag
\end{align}

By using this, Eq.~\eqref{eq2} can be written as
\begin{align}
    C^{(3)}&=\frac{\pi^3}{105}\int\frac{d^7k}{(2\pi)^7}\epsilon^{\mu\nu\rho\sigma\tau\lambda\delta}\tr G(\partial_{\mu}G^{-1})G(\partial_{\nu}G^{-1}) G(\partial_{\rho}G^{-1})G(\partial_{\sigma}G^{-1})G(\partial_{\tau}G^{-1})G(\partial_{\lambda}G^{-1})G(\partial_{\delta}G^{-1})\notag\\
    &=7\times\frac{\pi^3}{105}\int\frac{d^7k}{(2\pi)^7}\epsilon^{ijklmn}\tr G(\partial_{0}G^{-1})G(\partial_{i}G^{-1}) G(\partial_{j}G^{-1})G(\partial_{k}G^{-1})G(\partial_{l}G^{-1})G(\partial_{m}G^{-1})G(\partial_{n}G^{-1})\notag\\
    &=\frac{\pi^3}{15}\sum_{abcdef=G,E}\int\frac{d^7k}{(2\pi)^7}\epsilon^{ijklmn}\tr \frac{P_a(\partial_iP_G) P_b(\partial_jP_G)P_c(\partial_kP_G) P_d(\partial_lP_G) P_e(\partial_mP_G) P_f(\partial_nP_G)}{(i\omega-\epsilon_a)^2(i\omega-\epsilon_b)(i\omega-\epsilon_c)(i\omega-\epsilon_d)(i\omega-\epsilon_e)(i\omega -\epsilon_f)}(\epsilon_E-\epsilon_G)^6.   \label{eq10}
\end{align}
From the identities $P_G+P_E=1$ and $P_GP_E=P_EP_G=0$, we have
\begin{align}
    P_E\pdv{P_G}{k^i}=\pdv{P_G}{k^i}P_G, \quad
    P_G\pdv{P_G}{k^i}=\pdv{P_G}{k^i}P_E.\notag
\end{align}
Hence the trace in Eq.~\eqref{eq10} can be nonzero only when $(a,b,c,d,e,f)=(G,E,G,E,G,E)$ or $(E,G,E,G,E,G)$, giving
\begin{align}
    C^{(3)}&=\frac{\pi^3}{15}\int\frac{d^7k}{(2\pi)^7}\epsilon^{ijklmn}\tr \frac{P_G(\partial_iP_G) P_E(\partial_jP_G) P_G(\partial_kP_G) P_E(\partial_lP_G) P_G(\partial_mP_G) P_E(\partial_nP_G)}{(i\omega-\epsilon_G)^4(i\omega-\epsilon_E)^3}(\epsilon_E-\epsilon_G)^6\notag\\
    &+\tr \frac{P_E(\partial_iP_G) P_G(\partial_jP_G) P_E(\partial_kP_G) P_G(\partial_lP_G) P_E(\partial_mP_G) P_G(\partial_nP_G)}{(i\omega-\epsilon_G)^3(i\omega-\epsilon_E)^4}(\epsilon_E-\epsilon_G)^6\notag\\
    &=-\frac{\pi^3}{15}\int\frac{d^7k}{(2\pi)^7}\epsilon^{ijklmn}\frac{(\epsilon_E-\epsilon_G)^7}{(i\omega-\epsilon_G)^4(i\omega-\epsilon_E)^4}
    \tr (\partial_iP_G)P_E(\partial_jP_G) (\partial_kP_G)P_E(\partial_lP_G) (\partial_mP_G)P_E(\partial_nP_G) \notag\\
    &=-\frac{\pi^3}{15}\int_{-\infty}^{+\infty}\frac{id\omega}{2\pi} \int\frac{d\bm{k}}{(2\pi)^6}\epsilon^{ijklmn}\frac{(\epsilon_E-\epsilon_G)^7}{(i\omega-\epsilon_G)^4(i\omega-\epsilon_E)^4}
    \tr (\partial_iP_G)P_E(\partial_jP_G) (\partial_kP_G)P_E(\partial_lP_G) (\partial_mP_G)P_E(\partial_nP_G) \notag\\
    &=\frac{-i}{48\pi^3} \int d^6 \bm{k} \epsilon^{ijklmn} \tr (\partial_iP_G)P_E(\partial_jP_G) (\partial_kP_G)P_E(\partial_lP_G) (\partial_mP_G)P_E(\partial_nP_G).\label{eq11}
\end{align}

Finally, we write this equation in terms of the Berry curvature. Using the Berry connection,
\begin{equation}
    a_i^{\alpha\beta}=i\bra{\alpha\bm{k}}\pdv{k^i}\ket{\beta\bm{k}},\notag\\
\end{equation}
the Berry curvature is expressed by
\begin{align}
    f_{ij}^{\alpha\beta}&=\partial_ia_j^{\alpha\beta}-\partial_ja_i^{\alpha\beta}-i[a_i,a_j]^{\alpha\beta}\notag\\
    &=i\Bigl(\braket{\partial_i\alpha\bm{k}}{\partial_j\beta\bm{k}}-\braket{\partial_j\alpha\bm{k}}{\partial_i\beta\bm{k}}\Bigr)+i\Bigl(\braket{\alpha\bm{k}}{\partial_i\gamma\bm{k}}\braket{\gamma\bm{k}}{\partial_j\beta\bm{k}}-\braket{\alpha\bm{k}}{\partial_j\gamma\bm{k}}\braket{\gamma\bm{k}}{\partial_i\beta\bm{k}}\Bigr)\notag\\
    &=i\Bigl(\braket{\partial_i\alpha\bm{k}}{\partial_j\beta\bm{k}}-\braket{\partial_j\alpha\bm{k}}{\partial_i\beta\bm{k}}\Bigr)-i\Bigl(\braket{\partial_i\alpha\bm{k}}{\gamma\bm{k}}\braket{\gamma\bm{k}}{\partial_j\beta\bm{k}}-\braket{\partial_j\alpha\bm{k}}{\gamma\bm{k}}\braket{\gamma\bm{k}}{\partial_i\beta\bm{k}}\Bigr)\notag\\
    &=i\Bigl(\braket{\partial_i\alpha\bm{k}}{\partial_j\beta\bm{k}}-\braket{\partial_j\alpha\bm{k}}{\partial_i\beta\bm{k}}\Bigr)-i\Bigl(\bra{\partial_i\alpha\bm{k}}P_G\ket{\partial_j\beta\bm{k}}-\bra{\partial_j\alpha\bm{k}}P_G\ket{\partial_i\beta\bm{k}}\Bigr)\notag\\
    &=i\Bigl(\bra{\partial_i\alpha\bm{k}}P_E\ket{\partial_j\beta\bm{k}}-\bra{\partial_j\alpha\bm{k}}P_E\ket{\partial_i\beta\bm{k}}\Bigr).\notag
\end{align}
Thus we have
\begin{align}
    f_{ij}&=\sum_{\alpha\beta}\ket{\alpha \bm{k}}f_{ij}^{\alpha\beta}\bra{\beta\bm{k}}\notag\\
    &=i\Bigl((\partial_iP_G)P_E(\partial_jP_G)-(\partial_jP_G)P_E(\partial_iP_G)\Bigr). \notag
\end{align}
By using this, Eq.~\eqref{eq11} is transformed to
\begin{align}
    C^{(3)}&=\frac{1}{2^3}\times\frac{-i}{48\pi^3}\int d^6\bm{k}\epsilon^{ijklmn} \tr (-if_{ij})(-if_{kl})(-if_{mn})\notag\\
    &=\frac{1}{2^3}\times\frac{1}{48\pi^3} \int d^6\bm{k}\epsilon^{ijklmn} \tr f_{ij}f_{kl}f_{mn}\notag\\
    &=\frac{1}{48\pi^3}\int_{BZ} \tr f\wedge f\wedge f.\notag
\end{align}
\end{widetext}
\bibliography{reference}

\end{document}